\documentclass[aps,prl,superscriptaddress,twocolumn,showpacs]{revtex4-1}

\usepackage{amsmath, amsthm, amssymb}
\usepackage{graphicx}
\usepackage{color}
\usepackage{times}

\usepackage{natbib}
\usepackage{hyperref}
\hypersetup{
        colorlinks=true,
}

\usepackage[resetlabels,labeled]{multibib}
\newcites{supp}{References}

\newcommand{\nts}[1]{}

\newcommand{\eqnref}[1]{(\ref{#1})}

\long\def\beginpgfgraphicnamed#1#2\endpgfgraphicnamed{\includegraphics{#1}}

\begin{document}

\title{Gapped and gapless spin liquid phases on the Kagome lattice\\ from chiral three-spin interactions}

\author{Bela Bauer}
\affiliation{Station Q, Microsoft Research, Santa Barbara, CA 93106-6105, USA}

\author{Brendan P. Keller}
\affiliation{Physics Department, University of California,  Santa Barbara, CA 93106, USA}

\author{Michele Dolfi}
\affiliation{Theoretische Physik, ETH Zurich, 8093 Zurich, Switzerland}

\author{Simon Trebst}
\affiliation{Institute for Theoretical Physics, University of Cologne, 50937 Cologne, Germany}

\author{Andreas W. W. Ludwig}
\affiliation{Physics Department, University of California,  Santa Barbara, CA 93106, USA}

\begin{abstract}
We argue that a relatively simple model containing only SU(2)-invariant chiral three-spin interactions on a Kagome lattice of $S=1/2$ spins can give rise to both a gapped and a gapless quantum spin liquid. Our arguments are rooted in a formulation in terms of network models of edge states and are backed up by a careful numerical analysis.
For a uniform choice of chirality on the lattice, we realize the Kalmeyer-Laughlin state, i.e. a gapped spin liquid which is identified as the $\nu=1/2$ bosonic Laughlin state. For staggered chiralities, a gapless spin liquid emerges which exhibits gapless spin excitations along lines in momentum space, a feature that we probe by studying quasi-two-dimensional systems of finite width.
We thus provide a single, appealingly simple spin model (i) for  what is probably the simplest realization of the Kalmeyer-Laughlin state to date, 
as well as (ii) for a non-Fermi liquid state with lines of gapless SU(2) spin excitations.
\end{abstract}

\maketitle

Quantum spin liquids are elusive phases of matter where spins do not freeze into local ordering patterns even at zero temperature~\cite{balents2010}. Typically, such spin liquids come in two general flavors, namely gapped topological spin liquids and gapless spin liquids~\cite{wen2004}. Key properties of topological spin liquids are that they have ground-state degeneracies depending on the topology of the system and, when confined to two spatial dimensions, they host quasiparticle excitations with exotic anyonic particle exchange statistics. Many examples of such systems are known both with~\cite{Kitaev97,levin2005} and without~\cite{kalmeyer1987} time-reversal symmetry.
Systems in the other broad class, gapless spin liquids, are characterized by gapless excitations that are not Goldstone modes arising from spontaneous breaking of a symmetry. These states can be further divided into two groups: One where gapless excitations occur at singular points in momentum space -- such as, e.g., the case of algebraic spin liquids~\cite{hermele2005}, the celebrated Kitaev honeycomb model~\cite{kitaev2006} and variations thereof~\cite{yao2009}. In the other group, the gapless modes arise on surfaces in momentum space~\cite{paramekanti2002,motrunich2005,lee2005,motrunich2007,yao2009}, generalizing the notion of a Fermi surface to systems with bosonic character, frequently referred to as Bose surfaces~\cite{motrunich2007}. 
However, for the latter group of states, few results~\cite{motrunich2007,sheng2008,sheng2009,tay2010,tay2011,block2011,mishmash2011,jiang2012} firmly establish a connection between such states and microscopic models, or identify the fundamental theory characterizing the gapless modes on the Bose surface of such non-Fermi liquids.

Here, we explore the phases that emerge when promoting the scalar spin chirality
\begin{equation} \label{eqn:chi}
\chi_{ijk} = \vec{S}_i \cdot (\vec{S}_j \times \vec{S}_k),
\end{equation}
which is SU(2) invariant, but breaks both parity and time-reversal invariance, to an interaction between spins on a lattice built from triangles,
\begin{align} \label{eqn:ham}
H &= \sum_{i,j,k \in \bigtriangleup,\bigtriangledown} J_{ijk} \chi_{ijk} &J_{ijk} &= \pm1
\end{align}
where $i$, $j$, $k$ are always ordered clockwise around a triangle.
We show that depending on the assignment of the chiralities $J_{ijk}$ to the triangles of the lattice, 
this interaction
can give rise to either a chiral topological spin liquid or a gapless non-Fermi-liquid  state
with lines of gapless spin excitations in momentum space.

The existence of gapped chiral spin liquid phases, which are obtained as generalizations of fractional quantum Hall states to spins, was first hypothesized by Kalmeyer and Laughlin~\cite{kalmeyer1987}. Soon after this proposal, it was suggested that the scalar spin chirality~\eqnref{eqn:chi} could be relevant for stabilizing such a spin liquid phase~\cite{baskaran1989,wen1989}. However, making a stringent connection between a microscopic Hamiltonian and the sought-after chiral spin liquid state has remained elusive for many years. Some progress has recently been made in constructing somewhat artificial spin Hamiltonians~\cite{yao2007,schroter2007,thomale2009,greiter2012} involving decorated lattices or long-range interactions.
Furthermore, an alternative perspective arose from topological flat band models~\cite{tang2011,sun2011,neupert2011,wang2011}.

\begin{figure}[b]
  \beginpgfgraphicnamed{puddle1}
  \input{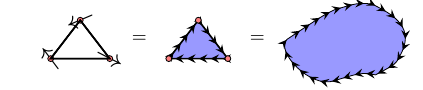}
  \endpgfgraphicnamed
  \caption{(Color online) Sketch of a puddle of topological phase replacing each triangle of three spins. \label{fig:puddle1} }
\end{figure}

\begin{figure}
  \beginpgfgraphicnamed{puddles}
  \input{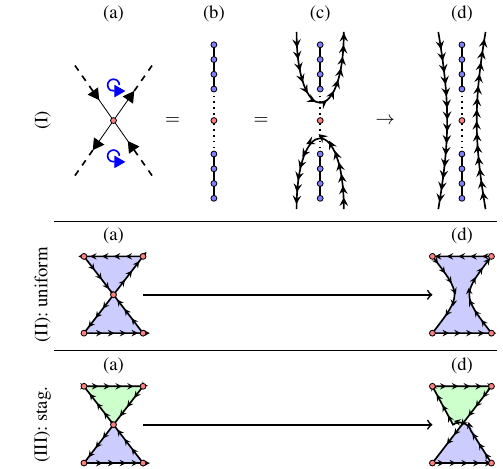}
  \endpgfgraphicnamed
  \caption{(Color online) Top panel: Illustration of the behavior of the edge states at a corner shared by two puddles. Bottom panel: Behavior of two corner-sharing triangular puddles of the topological phase. Different colors indicate different chiralities of the topological phases. \label{fig:puddle} }
\end{figure}

Here, we develop a powerful perspective rooted in the physics of network models of edge states akin to the Chalker-Coddington network model for the integer quantum Hall transition~\cite{chalker1988}. The key step is to view each triangle of spins as the seed of a chiral topological phase, a puddle encircled by an edge state, as illustrated in Fig.~\ref{fig:puddle1}. The natural candidate for the chiral topological phase filling the puddle is  the bosonic $\nu=1/2$ Laughlin state~\cite{halperin1983}\nocite{moore1991}, which is known to have the SU(2) symmetry required by our construction and
is
 also the state that was envisioned by Kalmeyer and Laughlin~\cite{kalmeyer1987}. Forming a lattice out of the elementary triangles, we should then consider a situation with many individual puddles of this topological phase. To see what collective state is formed, we have to understand how two puddles are joined. In doing so, it turns out to be useful to consider the specific situation of corner-sharing triangles, as realized for instance in a Kagome lattice.
This situation of edges meeting at the corner shared by two triangles is an incarnation of two-channel Kondo physics
(see e.g. Ref. \cite{AffleckLudwig1991,MaldacenaLudwig1997})): 
If we envision the puddles to be very large, they would carry the edge state on each side and the corner would look as shown in panel (I.a) of Fig.~\ref{fig:puddle}. The pair of edge states on the upper triangle is known~\cite{halperin1983,moore1991} to be described by the same theory as the right- and left-movers of a semi-infinite uniform spin-1/2 Heisenberg chain, and analogously for the lower pair of edge states. The spin at the corner then appears as the center spin of an infinite chain (panels (I.b,I.c)). It is well known in the context of two-channel Kondo physics that the infinite chain will heal if the center spin is coupled to the two semi-infinite chains with equal strength~\cite{eggert1992,kane1992}. Then, the right- and left-movers will extend throughout the entire, infinite system (panel (I.d)).
The effect on the corner spin is summarized in panel II of Fig.~\ref{fig:puddle}, where the situation shown in II.a corresponds to I.a, while II.d corresponds to I.d. As is evident from II.d, the corner spin has merged the two triangles to form a larger puddle encircled by a single edge state, i.e. to form a larger region of the topological phase.

\begin{figure}
  \beginpgfgraphicnamed{hom}
  \input{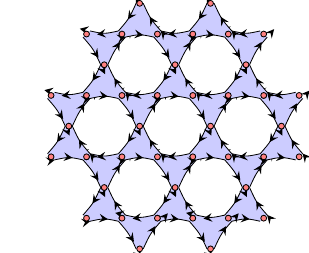}
  \endpgfgraphicnamed
  \caption{(Color online) Visualization of the topological phase for uniform choice of chiralities. A collective edge state encircles the whole systems, and closed edges encircle each hexagon.\label{fig:hom} }
\end{figure}

If we consider all triangles to have the \emph{same} chirality ($J_\bigtriangleup = J_\bigtriangledown$), we can repeat these steps for all corner-sharing triangles of the Kagome lattice. The system then forms one macroscopic, extended region of a \emph{single} topological phase with one edge state encircling its boundary, as illustrated in Fig.~\ref{fig:hom}, with closed loops encircling the interior hexagons of the Kagome lattice. We conclude that choosing uniform chiralities in our model, we obtain a direct realization of the Kalmeyer-Laughlin proposal for a chiral topological spin liquid phase.

\begin{figure}
  \begin{tabular}{c|c}
    \parbox{1.55in}{ \beginpgfgraphicnamed{stag} \input{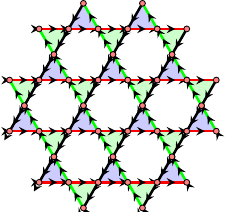} \endpgfgraphicnamed } &
    \parbox{1.9in}{ \includegraphics[width=1.9in]{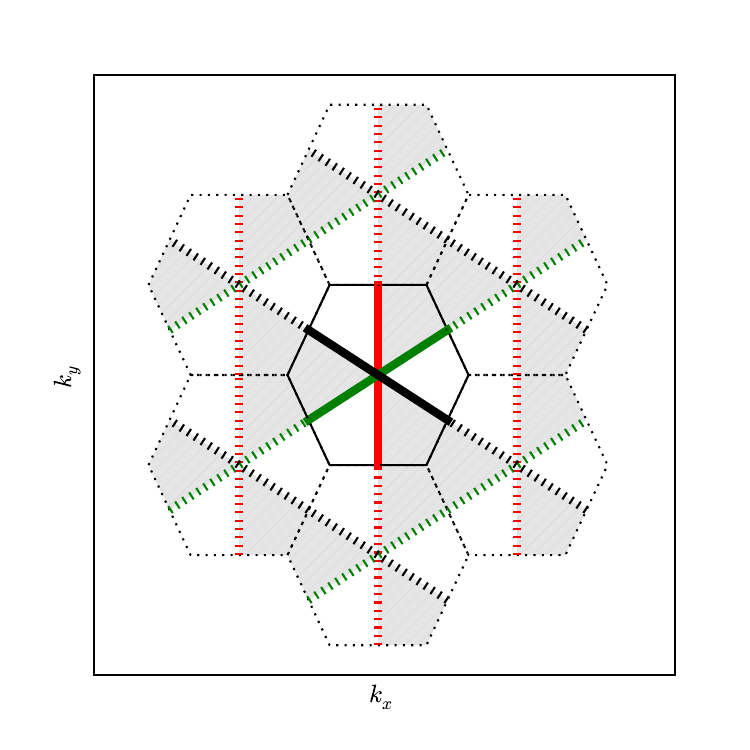} }
  \end{tabular}
  \caption{(Color online) Real-space (left) and momentum-space (right) picture of the staggered (gapless) phase. The colors correspond to the three independent directions of extended edge states. The inner hexagon, enclosed by a solid line, shows the first Brillouin zone. \label{fig:kagome2d} }
\end{figure}

It is now natural to ask whether other phases can be predicted from the same perspective when choosing different patterns of chiralities. It turns out that this is indeed the case for staggered chiralities, i.e. when the down-pointing and the up-pointing triangles of the Kagome lattice are assigned opposite chiralities ($J_\bigtriangleup = -J_\bigtriangledown$). For this pattern, we find the sought-after non-Fermi-liquid state with gapless spin excitations on lines in momentum space.

To see this, we again start from the case of two adjacent triangles, which are now filled with topological phases of opposite chirality (Fig.~\ref{fig:puddle} III.a). To see the effect on the edge states, consider changing the chirality of the upper triangle by exchanging its two top spins, that is twisting the upper triangle. The resulting pattern of edge states after healing is depicted in panel III.d; note that no backscattering occurs.
Again repeating this for all triangles of the Kagome lattice results in edge states extended throughout the entire system, as shown in the left
panel of Fig.~\ref{fig:kagome2d}. The pattern of edge states follows three sets of parallel lines rotated with respect to each other by 120 degree rotations.

To infer the shape of the resulting Bose surface of gapless excitations (corresponding to the extended edge states), we note that a single extended edge state, say in the $x$ direction, is gapless at a single point in momentum space, $k_x = 0$. Since all edges are decoupled, the gapless points corresponding to a stack of parallel edge states form a continuous line in momentum space perpendicular to the direction of $k_x$, i.e. at $k_x = 0$ and arbitrary $k_y$. This describes a line of gapless modes in momentum space. The full Bose surface for the Kagome lattice is then obtained when combining three such stacks of independent, gapless edge states, each rotated by 120 degrees with respect to each other. The resulting Bose surface, shown in the right panel of Fig.~\ref{fig:kagome2d}, is therefore comprised of three gapless lines~\footnote{The six-fold rotational symmetry of the underlying Kagome lattice is broken to a three-fold rotational symmetry, and the reflection symmetry $x \leftrightarrow -x$ has been broken, while $y \leftrightarrow -y$ remains a symmetry. The reflection $x \rightarrow -x$ combined with time reversal, however, is a symmetry of the system.}.

\noindent {\it Majorana model.--} Before discussing the technically challenging case of SU(2) spins, we illustrate the validity of the above network model picture in the situation where the spins are replaced by Majorana fermion zero modes. In particular, we can form a term analogous to the spin chirality~\eqnref{eqn:chi} which defines a notion of chirality for a triangle formed by sites $i$, $j$, $k$:
\begin{equation} \label{eqn:Hmaj}
\tilde{\chi}_{ijk} = \imath (\gamma_i \gamma_j + \gamma_j \gamma_k + \gamma_k \gamma_i).
\end{equation}
Here, $\gamma_i = \gamma_i^\dagger$ denote Majorana operators ($\gamma_i^2=1$, $\lbrace \gamma_i, \gamma_j \rbrace = \delta_{ij}$) on the sites of the Kagome lattice and the choice of sign for $J_{ijk}$ again sets the chirality for a triangle. This is a non-interacting system which can be solved exactly. 

Similarly to the situation of spins, we can view each triangle as the seed of a topological phase and apply arguments analogous to those discussed in the context of Fig.~\ref{fig:puddle}~\cite{kane1992}.
Validating our discussion in the context of Figs.~\ref{fig:hom} and \ref{fig:kagome2d}, we find that if the chiralities are chosen the same on all triangles, a gapped phase emerges (for details, refer to the supplemental material~\cite{supp}). Since it is a non-interacting fermion model, the Chern number $C$~\cite{thouless1982} is easily calculated~\cite{hastings2010,loring2010} and we find that $C=\pm 1$ depending on the overall chirality of the Hamiltonian. If, on the other hand, we choose staggered chiralities, we find a gapless system with a Fermi surface . The shape of the Fermi surface precisely matches the discussion above.\cite{supp} -  Related models have previously been considered~\cite{ludwig2011} on the triangular lattice~\cite{grosfeld2006,biswas2011} and Kagome lattice~\cite{shankar2009,chua2011} and both a topological state and a Fermi surface state have been obtained depending on whether the chiralities on all triangles are the same or staggered.

\noindent {\it Numerical methods.--} Returning to the original spin model~\eqnref{eqn:ham}, which is not exactly solvable, the nature of the phases is much harder to establish and one has to resort to large-scale numerical simulations. We briefly summarize our numerical approach for (a) the topological phase and (b) the gapless phase.

{\it (a) Topological phase}: To identify the topological nature of the phase, we rely on observing the ground-state degeneracy and the presence of edge states. Using exact diagonalization (ED) and the density-matrix renormalization group (DMRG)~\cite{white1992,white1992-1,schollwoeck2005,schollwoeck2011}, we show that there is a gap to all excitations and that when placed on a torus, there are two low-lying states in the $S_z=0$ sector, which are well-separated from the rest of the spectrum. We furthermore show that on a thin, long strip, the system displays a gapless edge state (see Fig.~\ref{fig:hom} and the supplemental material~\cite{supp}) which disappears when the long boundaries are connected to form a cylinder. All of these observations
provide  evidence for with a bosonic $\nu=1/2$ Laughlin state.

{\it (b) Gapless phase}: A characteristic of gapless phases with a Bose surface is that their quasi-one-dimensional precursors are critical with the number of gapless degrees of freedom growing with the width of the system~\cite{sheng2008,sheng2009}. We here focus on the particular case of a two-leg ladder, in which two gapless modes occur. We therefore expect to observe a central charge of $c=2$ (see supplemental material~\cite{supp}).  Using DMRG, we show that the system is indeed gapless (with the finite-size spin gap, $\Delta_s = E(S_z = 1) + E(S_z = -1)- 2 E(S_z = 0)$ vanishing as $N^{-1}$, where
$N$ is the total number of lattice sites) and confirm the expected  central charge using fits to the entanglement entropy (see supplemental material~\cite{supp}).

The DMRG method scales exponentially in the width of the system $W$, but only polynomially or linearly in the length $L$ of the system for critical and gapped systems, respectively. This allows us to approach thin, long systems, i.e. $W$ small and $L \rightarrow \infty$. It is based on a variational ansatz that can be systematically improved by increasing the number 
$M$ of states kept,
where the computational cost is known to grow as $\mathcal{O}(M^3)$. We fix $W=2$ and use up to $M=3600$ states.

\begin{figure}
  \begin{tabular}{cc}
    \includegraphics{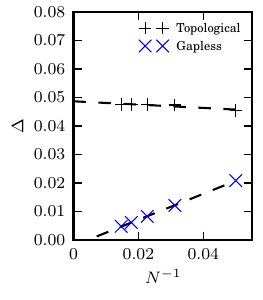}
    &\includegraphics{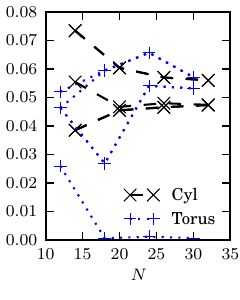} \\
     \multicolumn{2}{c}{\includegraphics{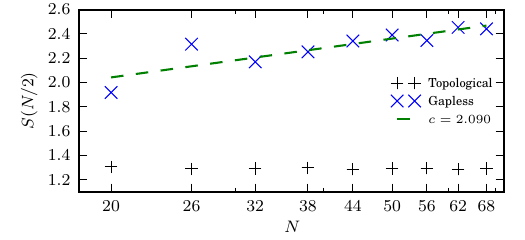}}
  \end{tabular}
  \caption{(Color online) In all panels, $N$ is the total number of lattice sites. Top left panel: Spin gap obtained by DMRG calculations on the cylinder. Top right panel: Exact diagonalization gaps of the uniform case on both the torus and the cylinder. Bottom panel: Finite-size entropy for cylinders. This plot analyzes the entropy at the center of the system (averaged over a number of sites to remove oscillations) as a function of the total system size. \label{fig:w2} }
\end{figure}

\begin{figure}
  \begin{tabular}{cc}
  \multicolumn{2}{c}{\includegraphics{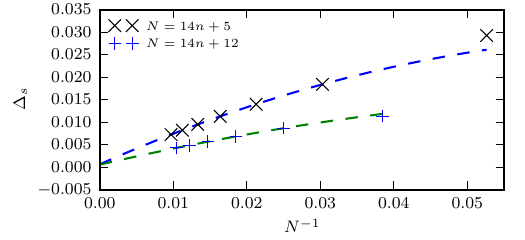}} \\
  \includegraphics{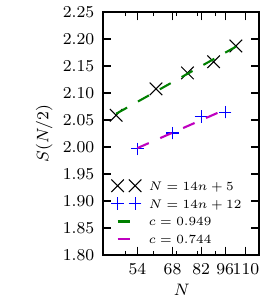} &\includegraphics{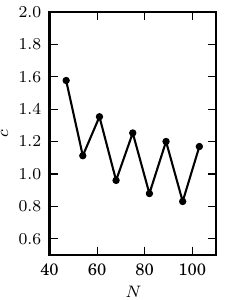}
  \end{tabular}
  \caption{(Color online) Top panel: Spin gap for the uniform phase on a thin, long strip. Bottom panels: Entanglement entropy for the uniform system on a strip. Bottom left: Entropy at the center of the system on a semi-logarithmic scale. Bottom right: Central charge extracted from a direct fit to the entropy for a given system size. Details of the fits can be found in the supplemental material~\cite{supp}. \label{fig:plane} }
\end{figure}

\noindent {\it Numerical results: (a) Topological phase.--} We start by discussing our results for the topological phase obtained for uniform chiralities. Our results for the finite-size spin gap $\Delta_s(N)$, where $N$ is the total number of sites, are shown in the upper left panel of Fig.~\ref{fig:w2}. It clearly approaches a finite value of $\Delta_s(\infty) \approx 0.05$. To further corroborate that the model has a finite correlation length, we can study the scaling of the entanglement entropy with system size (lower panel of Fig.~\ref{fig:w2}). We show clearly that the entanglement entropy at the center of the system becomes independent of the total system size even for relatively small systems, which is a clear signature of a gapped state. This is also consistent with the
behavior of correlation functions, which we find to decay exponentially (see supplemental material~\cite{supp}).

To identify the topological nature of this gapped phase, we study the dependence of the low-lying spectrum on the topology. We expect a two-fold degeneracy on the torus which is split by an exponentially small amount, and no degeneracy on the cylinder (i.e. a system with periodic boundary conditions in the transverse direction, but open boundary conditions in the longitudinal direction). The top right panel of Fig.~\ref{fig:w2} clearly reproduces this behavior: on the cylinder, we observe excited states close to the extrapolated spin gap $\Delta_s(\infty)$. On the torus, on the other hand, we find an additional low-energy state in the same $S_z$ sector very close to the ground state.

A final piece of evidence is obtained by placing the system on a fully open geometry, i.e. a strip. For this geometry, we expect a pair of extended edge states in the long direction to give rise to a gapless mode with central charge $c=1$. This is nicely confirmed by our results in Fig.~\ref{fig:plane}. In the top panel, we show the scaling of the spin gap with system size. While there is a strong even-odd effect, we clearly observe a scaling of $\Delta_s(N) \sim a/N + b/N^2$ (where $a$, $b$ are fit parameters) as expected for a critical system. To measure the central charge, we perform fits to the entanglement entropy, as shown in the bottom panels of Fig.~\ref{fig:plane}. We find that by either fitting the entropy at the center of the system as a function of system size, or by fitting the entropy as a function of block size for fixed system size, we obtain reasonable agreement with $c=1$.

\noindent {\it (b) Gapless phase.--} Turning to the gapless spin liquid phase for staggered chiralities, we again consider the scaling of the finite-size spin gap, shown in the top left panel of Fig.~\ref{fig:w2}. We find good agreement with $\Delta_s(N) \sim N^{-1}$, strongly indicative of a critical system. Determining the central charge of the critical theory for this gapless system, we find a fit to the entropy at the center of the system to be most reliable. Our data, which is shown in the lower panel of Fig.~\ref{fig:w2}, shows good agreement with the expected value $c=2$ of the central charge.

\noindent {\it Conclusions.--} Summarizing our results, we have obtained strong numerical evidence for (i) a gapped topological spin liquid and 
(ii) a gapless spin liquid with gapless spin excitations on lines in momentum space.
These phases are numerically observed in a relatively simple %(parameter-free) 
spin-1/2 model of three-spin interactions around the triangles of the Kagome lattice. The key concept to understand the emergence of these phases is a network model of edge states. For the case of the gapless spin liquid, the geometry of the gapless lines in momentum space 
follows directly from the aforementioned network model picture. 
The same reasoning,  which leads from the microscopic network model to the emergent 
nature of these phases,  can also be used and thereby confirmed
in a different, exactly solvable
model where the spins are replaced by Majorana fermion zero modes on the same lattice.

One aspect of our work that we plan to address in the future is the question of interactions between the gapless excitations on the lines in momentum space in the spirit of generalizations of Landau's Fermi liquid theory~\cite{haldane1994}. We also plan to explore the phase diagram when additional terms, such as a Heisenberg term on the bonds, are added to our model.

Future work should explore whether decoupling the Heisenberg antiferromagnet on the Kagome lattice using the well-known relation $\chi_{ijk}^2 \sim (S_i + S_j + S_k)^2$~\cite{wen1989} can lead to new insights into the much debated nature of the ground state of this model. Another interesting future direction to pursue is to make a connection to orbital currents induced by finite spin chiralities, as argued in Ref.~\cite{bulaevskii2008}.

\acknowledgements
\noindent {\it Acknowledgements.--}
The DMRG code was developed with support from the Swiss Platform for High-Performance and High-Productivity Computing (HP2C) and based on the ALPS libraries~\cite{bauer2011-alps}. S.T. was supported, in part, by  SFB TR 12 of the DFG. A.W.W.L was supported, in part, by NSF DMR-0706140. We acknowledge illuminating discussions with participants of the KITP workshop \emph{Frustrated Magnetism and Quantum Spin Liquids} (Fall 2012).

\bibliographystyle{apsrev4-1}
\bibliography{refs}

\clearpage

\section{Supplementary material}

This supplementary material is divided into three sections: in Section A., we discuss in more detail how the gapless lines in momentum space are obtained from the network model of edge states for the model with staggered chiralities (gapless spin liquid). In Section B., we explain in detail how the quasi-one-dimensional precursor systems of the two-dimensional gapped and gapless phases can be understood in terms of the network picture. In Section C., we focus on the special case of the (exactly solvable) Majorana hopping problem, which is obtained when the spins of the original model are replaced by Majorana zero modes, and we give some detail on its diagonalization. In particular, we calculate the shape of the Fermi surface explicitly for this model.
In Section D. we briefly discuss entanglement entropy and central charge, whereas in Section E. we discuss the exponential decay
of spin correlation functions in the gapped, topological phase.

\subsection{A. Gapless lines in momentum space}

\begin{figure*}
  \includegraphics[width=7in]{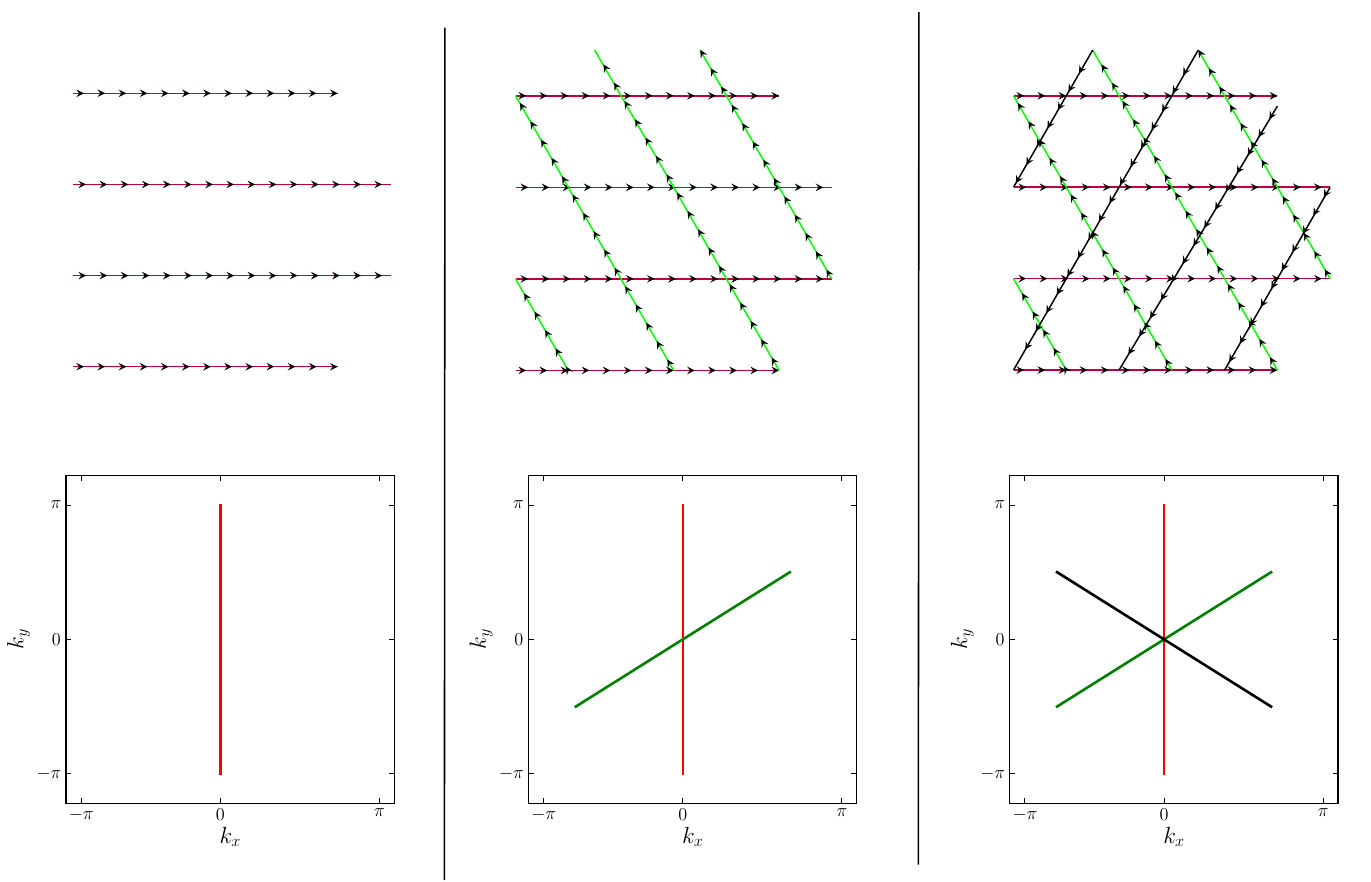}
  \caption{(Color online) This figure illustrates the construction of the gapless lines in momentum space (lower panels) from the real-space picture (upper panels). Each of the three sets of parallel lines forming the Kagome lattice gives rise to a line of gapless points in momentum space. The horizontal lines give rise to gapless states at $k_x=0$ (left panel). The two other lines are rotated by 120 and 240 degrees, respectively. \label{fig:fs} }
\end{figure*}

In Figure~\ref{fig:fs}, we show in more detail how to infer the location where gapless states exist in momentum space (i.e, Fermi surface in the case of fermions) from the network model of edge states in real space. The top left panel of Fig.~\ref{fig:fs} shows a stack (in the $y$ direction) of uncoupled edge states directed in the $x$ direction. The resulting picture in momentum space is shown in the lower left panel of the figure: An edge state in the (say)  
$x$ direction is gapless at one point in momentum space, i.e.  when the momentum $k_x$ vanishes, $k_x = 0$. Since the different edges are decoupled, the gapless points of the stack of parallel edge states form a continuous line in momentum space perpendicular to the direction of the edges, i.e. at $k_x = 0$ and arbitrary $k_y \in [-\pi,\pi)$. This describes a line of gapless modes in momentum space.

The full real-space network on the Kagome lattice, which is shown in Fig.~\ref{fig:kagome2d} of the main paper, is recovered when three such stacks of gapless edges,  rotated by 120 degrees with respect to each other, are combined.
 In the center panel of Fig.~\ref{fig:fs}, we show the situation for two such stacks, i.e. a second stack of such edges, rotated by 120 degrees with respect to the edges in $x$ direction, has been added. This additional stack of edges will give rise to an additional line of gapless states in momentum space, which is rotated by 120 degrees with respect to the first one as well, as shown in the lower center panel of Fig.~\ref{fig:fs}. In the right-most panel, all three stacks are combined, giving rise to three lines of gapless states in momentum space rotated with respect to each other by 120 degrees.

\subsection{B. Quasi-one-dimensional precursors of the two-dimensional phases}

In this section, we provide additional information on the quasi-one-dimensional precursors of the two-dimensional phases. Since the DMRG method scales exponentially in the width of the system while scaling only polynomially in its length, we perform our numerical simulations on quasi-one-dimensional systems.

\begin{figure*}[p]
  \beginpgfgraphicnamed{kagome2o}
  \input{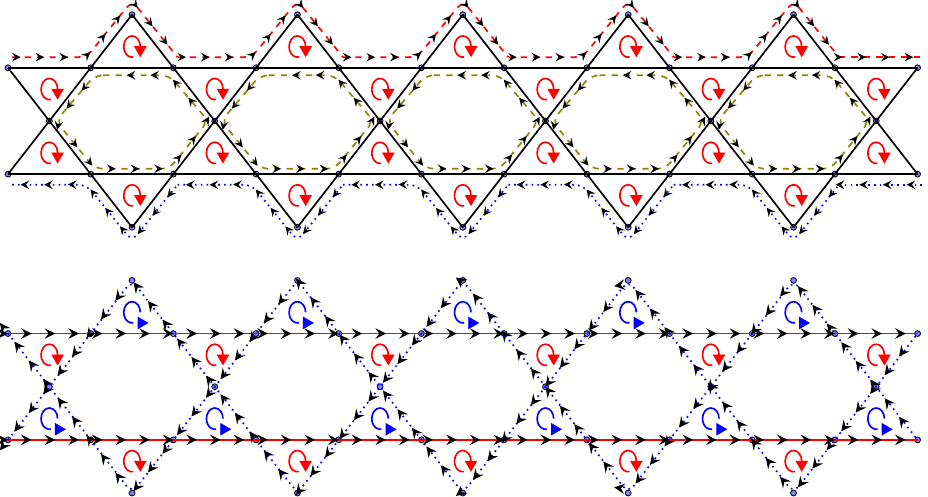}
  \endpgfgraphicnamed
  \caption{(Color online) Two-leg Kagome systems with illustration of the choice of chiralities for the uniform and staggered model, in quasi-one-dimensional
stip geometry. The red dashed lines indicate right-moving extended states, blue dashed line indicate left-moving extended states, and the green dashed lines encircling the hexagon show the localized states in the topological phase.\label{fig:kagome2o} }
\end{figure*}

Figure~\ref{fig:kagome2o} shows a quasi-one-dimensional version of the networks discussed in the main text. The top panel
refers to the case where the corresponding two-dimensional system, depicted in  Fig.~\ref{fig:hom} of the main text, is
the gapped 2D phase.
The bottom panel refers
to the case where the corresponding two-dimensional system, depicted in  Fig.~\ref{fig:kagome2d} of the main text, 
is the gapless 2D phase.
In the figure, we show a system with open boundary conditions in the transverse direction, and with a width of $W=2$ and a total number of sites $N=40$.
Periodic boundary conditions in the transverse direction are obtained by identifying the top-most spin of the top row of (up-pointing) triangles with the bottom-most spin of the bottom row of (down-pointing) triangles. In this case, the topology is that of a surface of a cylinder or torus, for open or periodic boundary conditions in the longitudinal direction, respectively.

The upper panel of Fig.~\ref{fig:kagome2o} shows the uniform case, which is predicted to be in a topological  gapped chiral spin liquid phase. In the bulk, this phase has no extended edges; instead, all edges combine to form localized states encircling the hexagons. These are shown as green dashed lines in the figure. For the open boundary conditions shown in the figure, extended edge states form along the boundaries of the system in the long (horizontal) direction. These edge states are shown as red dashed and blue dotted lines in the figure. From this picture, we predict that the central charge for this quasi-one-dimensional system is $c=1$ (corresponding to the edge of a single bosonic $\nu=1/2$ Laughlin state described by an SU$(2)_1$ CFT), which is confirmed by the numerical calculation of the entanglement entropy, which is detailed in the main paper.

The lower panel of Fig.~\ref{fig:kagome2o}, on the other hand, shows the staggered case, which is predicted to be a gapless spin liquid phase. For the quasi-one-dimensional system shown in the figure, one set of extended edges forms along the horizontal lines of the Kagome lattice, shown here as solid red line. These are right-moving. While in the truly two-dimensional system, two additional, independent stacks of extended edges would appear on the two other sets of lines which are rotated by 120 and 240 degrees, the quasi-1d geometry forces these edges to follow the zig-zag pattern shown as dotted blue lines in the figure. Those edges are left-moving. Counting the total number of edges shows that there are two left-moving and two right-moving edges leading a the central charge of $c=2$.

\begin{figure}
  \includegraphics[width=7cm]{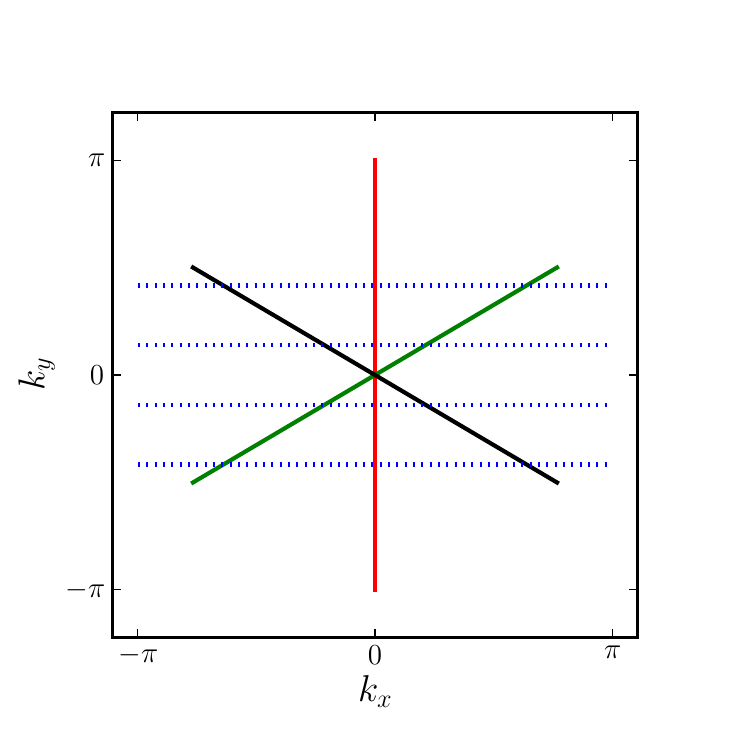}
  \caption{(Color online) Gapless lines (Fermi surface) of the two-dimensional model, cf. Fig.~\ref{fig:kagome2d} of the main text. In this figure, we additionally sketch a set of allowed momenta (dashed blue lines) which are obtained if the system is placed on a quasi-one-dimensional geometry. \label{fig:fssbs} }
\end{figure}

Figure~\ref{fig:fssbs} shows how the results for the quasi-one-dimensional systems discussed in the previous paragraph indicate the existence of extended lines of gapless modes in momentum space in the two-dimensional gapless phase. Making the system finite in the transverse direction leads to discretized momenta in this direction, where the number of such allowed momenta grows linearly in the width of the system. In Figure~\ref{fig:fssbs}, these allowed momenta are indicated as blue dotted horizontal  lines. The system is now gapless only at a finite number of points in momentum space which are given by the intersection of the allowed momenta with the gapless lines of the fully two-dimensional system. For such a gapless phase, the central charge measured in a quasi-one-dimensional geometry grows linearly in the number of such intersections.

\subsection{C. Non-interacting Majorana model}

\begin{figure}
  \centering
  \includegraphics[width=6cm]{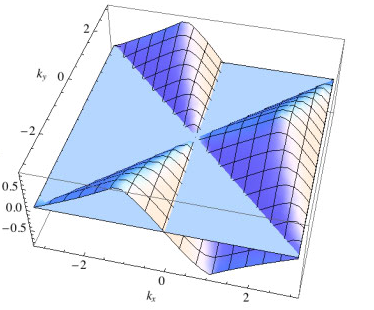}
  \caption{(Color online) Intersection of the central band $E_2(k_x,k_y)$ of the staggered model with the $E=0$ plane (light blue). The lines where the $E=0$ plane crosses the band represent the gapless lines; areas with $E_2(k_x,k_y) < 0$ correspond to filled regions of momentum space. \label{fig:intersect}}
\end{figure}

In the following section, we will discuss some details of the band structure obtained for the Majorana hopping Hamiltonian (cf. Eqn.~\eqnref{eqn:ham})
\begin{align}
H &= \sum_{\bigtriangleup} \chi_{ijk} \pm \sum_{\bigtriangledown} \chi_{ijk} \\
\chi_{ijk} &= \imath (\gamma_i \gamma_j + \gamma_j \gamma_k + \gamma_k \gamma_i),
\end{align}
where the case with the $+$ ($-$) sign corresponds to the uniform (staggered) phase. In both cases, the Hamiltonian is quadratic in the Majorana fermion operators and can therefore be diagonalized straightforwardly.
Since the Kagome lattice is obtained as a triangular lattice of three-site unit cells, we obtain three energy bands $E_\alpha(k_x,k_y)$, $\alpha = 1,2,3$.

In the uniform phase, we observe that all three bands are separated by a gap from each other, and the central band $E_2$ is dispersionless, i.e. $E_2(k_x,k_y) = 0$. Noting that the number of states in this band coincides with the number of hexagons in the system, we identify these zero-energy states as the non-interacting edge states encircling the hexagons of the Kagome lattice, which are shown in Fig.~\ref{fig:hom} of the main text. We calculate the Chern number using the prescription of Refs.~\cite{hastings2010,loring2010} and find that the Chern number $C$ of the top and bottom band is $C=\pm 1$, i.e. that the model is in a topological phase.

\begin{figure*}[p]
  \centering
  \includegraphics[width=5cm]{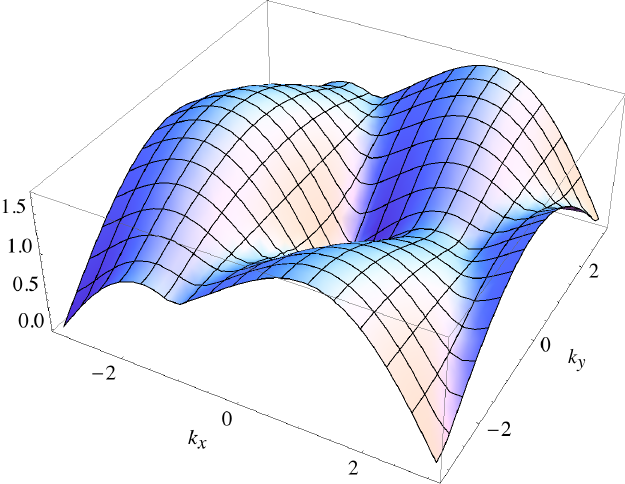}
  \includegraphics[width=5cm]{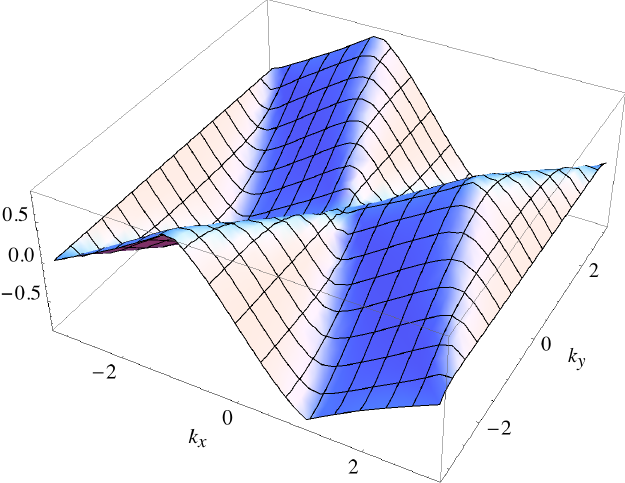}
  \includegraphics[width=5cm]{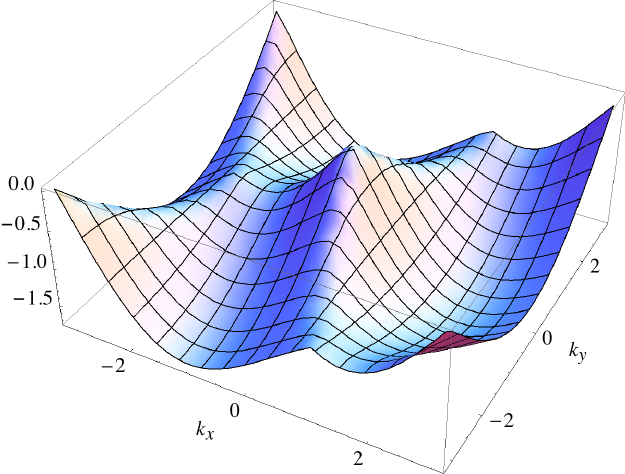}
  \caption{(Color online) The three bands $E_\alpha(k_x,k_y)$ of the Majorana hopping problem. Left panel: top band $E_3 \geq 0$, center panel: middle band $E_2$, right panel: bottom panel $E_1 \leq 0$. \label{fig:3bands} }
\end{figure*}

Our results for the staggered phase are summarized in Figures~\ref{fig:3bands} and \ref{fig:intersect}. Fig.~\ref{fig:3bands} shows all three bands. The top band has $E_3(k_x,k_y) \geq 0$, while the bottom band has $E_1(k_x,k_y) \leq 0$. All three bands are gapless and touch at $k_x = k_y = 0$. In Fig.~\ref{fig:intersect}, we show the intersection of the central band $E_2$ with the $E=0$ plane. This clearly shows the three straight lines where this band becomes gapless, as well as the regions of momentum space of filled and empty states. The agreement between this and the picture developed in the main text confirms that our reasoning, which is based purely on considering a network model of edge states, is correct.

\subsection{D. Entanglement entropy and the central charge}

From the ground states obtained by DMRG, the entanglement entropy for a contiguous block of $n$ sites at the end of an open system of $N$ total sites can be extracted without additional cost. This allows us to extract the central charge by performing a fit to the well-known result~\cite{holzhey1994,calabrese2004} for the entanglement entropy of a contiguous block of sites in an open $1d$ system,
\begin{equation} \label{eqn:S}
S(n) = S_0 + \frac{c}{6} \log \left( \frac{2N}{\pi} \sin \frac{\pi n}{N} \right),
\end{equation}
which for $n=N/2$ reduces to
\begin{equation} \label{eqn:Sc}
S(n=N/2) = S_0 + \frac{c}{6} \log \left( \frac{2N} {\pi} \right).
\end{equation}
Using Eqn.~\eqnref{eqn:S}, the central charge can be obtained from the ground state for a given system size $N$ by fitting to the entanglement entropy for various subsystem sizes $n$. In most cases, this is the most efficient way to extract the central charge. In some cases, however, it is more reliable to use Eqn.~\eqnref{eqn:Sc}, which requires the ground state for several different total system sizes $N$.

\subsection{E. Exponential decay of correlation functions in the topological phase}

\begin{figure}
  \includegraphics{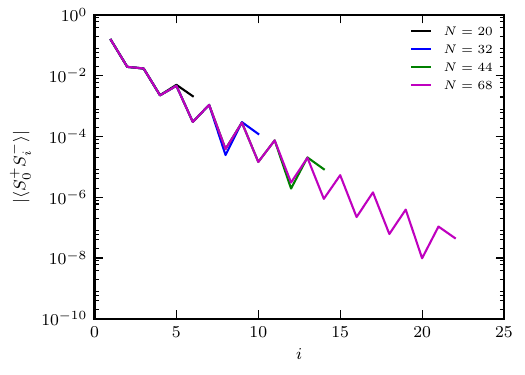}
  \caption{(Color online) Correlation function in the topological phase for a cylindrical topology with $W=2$. The correlation function shown is measured along one of the two horizontal chains that comprise the system, and $i$ denotes the distance from the end of the system along the chain. \label{fig:corr} }
\end{figure}

In Fig.~\ref{fig:corr}, we show a correlation function in the topological phase. We have measured $\langle S_0^+ S_i^- \rangle$, where all sites are chosen to be on one of the horizontal chains that comprise this system, and $i$ measures the distance along this chain. We find very good agreement with an exponential decay, i.e.
\begin{equation}
| \langle S_0^+ S_i^- \rangle | \sim \exp{ \left( -i / \xi \right) }
\end{equation}
with $\xi \approx 0.44$. This exponential decay is consistent with a gapped, topological phase.

\end{document}